\documentclass[twocolumn,tighten]{aastex63}

\newcommand{\halpha}{H$\alpha$}

\newcommand{\mdot}{$\dot{\text{M}}$}
\newcommand{\wten}{W$_{10}$}

\def\msunyr{\rm{M_{\sun} \, yr^{-1}}}
\def\mjupyr{\rm{M_{Jup} \, yr^{-1}}}
\def\kms{\rm \, km \, s^{-1}}
\newcommand{\ri}{R$_{\rm i}$}
\newcommand{\rw}{W$_{\rm r}$}
\newcommand{\tmax}{T$_{\rm max}$}
\newcommand{\cosi}{$\cos(i)$}
\newcommand{\lhalpha}{L$_{\rm H\alpha}$}

\received{---- --, ----}
\revised{---- --, ----}
\accepted{---- --, ----}
\submitjournal{ApJ}

\shorttitle{Magnetospheric Accretion of Planets around PDS 70}
\shortauthors{Thanathibodee et al.}

\begin{document}

\title{Magnetospheric Accretion as a Source of H$\alpha$ Emission from Proto-planets around PDS 70}

\correspondingauthor{Thanawuth Thanathibodee}
\email{thanathi@umich.edu}

\author[0000-0003-4507-1710]{Thanawuth Thanathibodee}
\affiliation{Department of Astronomy, University of Michigan, 323 West Hall, 1085 South University Avenue, Ann Arbor, MI 48109, USA}

\author[0000-0002-3950-5386]{Nuria Calvet}
\affiliation{Department of Astronomy, University of Michigan, 323 West Hall, 1085 South University Avenue, Ann Arbor, MI 48109, USA}

\author[0000-0001-7258-770X]{Jaehan Bae}
\affiliation{Department of Terrestrial Magnetism, Carnegie Institution for Science, 5241 Broad Branch Road NW, Washington, DC 20015, USA}

\author{James Muzerolle}
\affiliation{Space Telescope Science Institute, 3700 San Martin Drive, Baltimore, MD 21218, USA}

\author[0000-0002-1650-3740]{Ramiro Franco Hern\'andez}
\affiliation{Instituto de Astronom\'ia y Meteorolog\'ia, Universidad de Guadalajara, Avenida Vallarta No. 2602, Col. Arcos Vallarta, CP 44130, Guadalajara, Jalisco, M\'exico}

\begin{abstract}
Advances in high-resolution imaging have revealed
H$\alpha$ emission separated from the host star.
It is generally believed that the emission is associated with forming planets in protoplanetary disks. However, the nature of this emission is still not fully understood.
Here we report a modeling effort of H$\alpha$ emission from the planets around the young star PDS~70. Using standard magnetospheric accretion models previously applied to accreting young stars, we find that the observed line fluxes can be reproduced using a range of parameters relevant to PDS~70b and c, with the mean mass accretion rate of log(\mdot) = $-8.0\pm0.6$\, M$_{\rm Jup}$\,yr$^{-1}$ and $-8.1\pm0.6$\, M$_{\rm Jup}$\,yr$^{-1}$ for PDS~70b and PDS~70c, respectively.
Our results suggest that H$\alpha$ emission from young planets can originate
in the magnetospheric accretion of mass from the circumplanetary disk. We find that empirical relationships between mass accretion rate and H$\alpha$ line properties frequently used in T Tauri stars are not applicable in the planetary mass regime. In particular, the correlations between line flux and mass accretion rate underpredict the accretion rate by about an order of magnitude, and the width at the 10\% height of the line is insensitive to the accretion rate at {\mdot} $< 10^{-8}$\, M$_{\rm Jup}$\,yr$^{-1}$.
\end{abstract}


\section{Introduction} \label{sec:intro}

One of the thrilling new frontiers in exoplanet research is the unraveling of the earliest stages of planet formation. 
The advances of high-contrast, high-spatial-resolution technology have produced direct and indirect evidence of planets forming in circumstellar disks since very early ages. The regions cleared of small dust in the transitional disks \citep[c.f.][]{espaillat2014} were the first structures identified in protoplanetary disks, and they were attributed to the interaction of forming planets with the disk. 
Submillimeter interferometry has now shown that structures are present in probably most disks, and in particular, ringed structures are ubiquitous in disks around stars of all spectral types \citep{andrews2018}. Again, these structures are thought to be due to planets forming in disks \citep[e.g.,][]{bae2017,zhu2011}, although alternative hypotheses such as dust accumulation in condensation fronts of different molecular species \citep{zhang2015} have also been discussed. The best evidence of young planets comes from direct imaging, and so far detections have been claimed for a handful of systems with different degrees of reliability. For example, features in {\halpha} image around LkCa~15 are thought to be from planets \citep{kraus2012,sallum2015}, but later studies have shown that such features could come from the circumstellar disk \citep{mendigutia2018,currie2019}. 
\citet{reggiani2018} reported a detection of a planet candidate interior of the spiral arm of the disk around MWC 758; however, \citet{wagner2019} could not confirm this detection and instead found another candidate exterior of the spiral arm.
Other claims of planets around young stars include 
HD 100546 \citep{quanz2013a,quanz2015,brittain2014,currie2015,follette2017},
HD 169142 \citep{quanz2013b,biller2014,reggiani2014},
HD 163296 \citep{pinte2018,teague2018a}.

Robust planet detections have been reported
around the star PDS 70,
a K7 star in the $\sim$ 5-10 Myr old Upper Sco association.
PDS 70 is surrounded by a disk with an $\sim$ 80\,au cavity \citep{riaud2006,hashimoto2012,hashimoto2015,keppler2018}; 
it is also a pre-transitional disk \citep{espaillat2007}, that is, a disk with a large cavity but with another optically thick disk in the innermost au from the star. 
This configuration was first identified by SED fitting \citep{hashimoto2012,dong2012},
and later confirmed by observations in the NIR with SPHERE \citep{keppler2018} and in the submillimeter with ALMA \citep{long2018b,keppler2019}. 
\citet{keppler2018} reported the discovery of a companion at $\sim 22$ au from the star in the gap between these two disks,
confirmed by the 4-$\sigma$ detection of H$\alpha$ emission at the location of 
the companion using MagAO narrow filters \citep{wagner2018}.
The companion, PDS 70b, has a mass between 5 and 14 Jupiter masses, as indicated by a comparison of magnitudes and colors with different model predictions. 
However, hydrodynamic simulations including PDS 70b on a circular orbit failed to reproduce the large width of the cavity, and an additional companion beyond the orbit of PDS 70b was suggested \citep{keppler2019}. Indeed, using adaptive-optics-assisted integral-field spectroscopy with VLT/MUSE, \citet{haffert2019} detected {\halpha} emission from PDS~70b and reported additional {\halpha} emission from a second planet, PDS~70c.

High angular resolution observations by ALMA revealed the presence of 855\,$\mu$m dust emission from a circumplanetary disk (CPD) at the position of PDS~70c, with an estimate of dust mass between $2-4.2\times10^{-3}$\,M$_{\earth}$ \citep{isella2019}. CPD detection around PDS~70b is still unclear since ALMA observation \citep{isella2019} shows a slight offset between the position of the planet and that of the 855\,$\mu$m emission. Nevertheless, an estimated dust mass of $1.8-3.2\times10^{-3}$\,M$_{\earth}$ could be attributed to CPD of PDS~70b if it is present.

Mass accretion rates have been estimated for both PDS~70b and c. 
\citet{wagner2018}
detection of H$\alpha$ emission at the location of PDS~70b
led them to conclude
that the planet is actively accreting mass from its disk. 
They used the
H$\alpha$ flux 
to estimate a mass accretion rate onto the planet of
\mdot $\sim 10^{-8} M_{\rm Jup}\,{\rm yr}^{-1}$, which is consistent with the upper limit estimated by
\citet{christiaens2019} from the lack of detection in Br$\gamma$. 
\citet{haffert2019} 
also inferred values of the mass accretion rate,
but they used
the width at the 10\% height  (\wten)
of the {\halpha} line estimates for PDS~70b and c,
applying the correlation from \citet{natta2004}. 
The mass accretion rates for PDS~70b determined by \citet{haffert2019} is similar to that by \citet{wagner2018}, even though the {\halpha} flux is an order of magnitude lower. This is due to 
uncertainties in the diagnostics 
used to estimate mass accretion rates.

The correlations used to estimate mass accretion rates from emission line properties are determined from line luminosities and accretion luminosities
calculated for T Tauri stars \citep[e.g.,][]{natta2004,rigliaco2012,ingleby2013}.
Since accretion in that type of stars occurs through a magnetosphere, the 
assumption that the T Tauri star L$_{\rm line}$ vs. L$_{\rm acc}$ calibrations 
are valid for planets 
would imply that mass is loaded from the CPD onto the planet through magnetospheric accretion as well. However, this has not yet been proven.

The relationship between {\halpha} line luminosity and accretion properties
have also been 
explored to some extent in other types of models. It has been suggested that {\halpha} emission arises in regions of the order of a few hundred planet radii heated to ${\rm 8000 - 10000\,K}$ by the shock on the CPD surface due to infalling gas from the circumstellar disk \citep{szulagyi2017b}. Using 1D radiative transfer model, \citet{aoyama2018} calculated the properties of the CPD shock and predicted the {\halpha} fluxes for different densities and planetary masses for an application to the suggested planet around LkCa~15. Nevertheless, given the low surface density of the circumstellar disk around PDS~70 at the locations of both planets \citep[$\sim10^{-2}$g\,cm$^{-2}$;][]{keppler2019}, the model of \citet{aoyama2018} would imply a weak {\halpha} emission, more than an order of magnitude lower than observed in PDS~70b and c.

In this paper, we explore the possibility 
that the H$\alpha$ flux associated with the
young planets in PDS 70 system arises in magnetospheric flows akin to those present
in young stars surrounded by accretion disks. According
to the magnetospheric model for
accretion, the stellar magnetic field truncates the disk and matter falls towards the star at nearly free-fall velocities, merging with the photosphere through a shock at the stellar surface
\citep{hartmann2016}. For typical magnetic field strengths and mass accretion rates, the truncation radius is of the order of a few stellar radii. 
The line profiles and strength of the emission lines seen in the optical and near-IR spectra of young, low mass stars, i.e., T Tauri stars, can be explained by formation on the magnetospheric accretion flows \citep{muzerolle1998a,muzerolle2001,thanathibodee2019}. For this model to be valid for planets, a planetary magnetic field with sufficient strength
to truncate the disk at several planetary radii is needed,
as well as densities and temperatures in the magnetospheric flows 
capable of producing an H$\alpha$ line comparable with observations. 

The core accretion mechanism postulates three main phases in the formation process of giant planets: accumulation of planetesimals and/or pebbles into a core, slow gas accretion in an envelope, and runaway growth of the gas envelope, when the mass of the envelope reaches the mass of the core \citep{pollack1996}. At or near the last stages of the runaway phase, 
mass is transferred from the circumstellar disk to the
CPD and from it to the planet \citep{papaloizou2005,ginzburg2019}. Detailed Hydrodynamic (HD) and Magnetohydrodynamic (MHD) numerical simulations of the flows inside the Hill radius
show that mass flows onto the CPD from high latitudes, almost vertically, and that some material on the CPD midplane actually flows back into the circumstellar disk, away from the system \citep{machida2008,tanigawa2012,gressel2013,szulagyi2016,szulagyi2019}. 
The mass flows are not axially symmetric and are mostly directed toward the planet only well inside a few percents of the Hill radius, as the high-resolution simulations of \citet{tanigawa2012} show. For parameters corresponding to Jupiter, the Hill radius is $R_H \sim$ 0.34 au $\sim$ 730 Jupiter radii. Since these simulations mainly focus on the disks, their resolutions 
do not reach regions much closer to the planet, and 
they do not show how mass actually transfers onto the planetary core.

The region of interaction between a planet and its CPD has been explored by \citet{batygin2018}. This author proposes that the interior of the forming planet becomes convective because of its high luminosity, and a magnetic field of the order of 0.5 - 1 kG  is generated as a result. This field truncates the disk at a few, 4 - 5, planet radii,  and matter accretes onto the planet following the field lines. As noted by \citet{batygin2018}, this picture is similar to that of a magnetosphere around young stars, with the difference that in the latter, the mass comes from the circumstellar disk. In the case of planets, matter coming from the circumstellar disks and falling vertically onto the CPD is deflected towards the planet as it reaches the field line that has truncated the CPD. The rest of the infalling matter falls onto the CPD and is eventually expelled back towards the circumstellar disk \citep{batygin2018}.

In this paper, we use planetary parameters inferred from observations of PDS~70b and c and calculate the H$\alpha$ fluxes and line profiles for a range of suitable mass accretion rates and temperatures.
In \S~\ref{sec:acc_model}, we describe the magnetospheric model and in \S~\ref{sec:results}, the results of our exploration. Finally, we give a brief discussion and conclusions in \S~\ref{sec:discussion}.


\section{Magnetospheric Accretion Model} \label{sec:acc_model}
\subsection{Model Description}
To calculate the {\halpha} line flux formed in a magnetosphere around a planet, we use the magnetospheric accretion model \citep{muzerolle2001} initially developed for T Tauri stars. The detailed description of the model can be found in \citet{hartmann1994,muzerolle1998a,muzerolle2001}. Here we describe the main assumptions and parameters.

This model assumes an axisymmetric accretion flow arising from a co-rotating Keplerian gas disk on the same plane as the planet's equator. The material flows in the magnetic dipole geometry characterized by the truncation radius (R$_{\rm i}$) and the width at the based of the flow (W$_{\rm r}$). The density at a given point in the flow
follows a steady flow prescription, parameterized by the total mass accretion rate {\mdot}.
At each point in the flow, the temperature scales with the density; the maximum temperature T$_{\rm max}$, a free parameter, describes each model.
The model uses a 16-level hydrogen atom and the mean intensities of each transition and level populations are calculated by adopting the extended Sobolev approximation. 
The line specific intensity is calculated using a ray-by-ray method for a given inclination angle ($i$) between the line of sight and the planet rotation/magnetic axis. The line flux density is calculated by spatially integrating the intensity. The final, continuum-subtracted line flux is calculated over the range of $\pm$500\,$\kms$ since the relevant range of velocities would be on the order of the planets' escape velocity $\sim130\,\kms$ for the planets' parameters.

\subsection{Grids of Models}

\begin{deluxetable}{lccc}[t]
\tablecaption{Range of Model Parameters \label{tab:model_param}}
\tablehead{
\colhead{Parameters} & \colhead{Min.} & \colhead{Max.} & \colhead{Step} 
}
\startdata
log({\mdot/$\mjupyr$})	    & -11	& -5 	& 1  \\
T$_{\rm max}$ (K)	        & 8000	& 11000	& 200 \\
R$_{\rm i}$	(R$_{\rm p}$)	& 2	    & 8	    & 2 \\
W$_{\rm r}$ (R$_{\rm p}$)	& \multicolumn{3}{c}{1, 2, 4, 6} \\
cos($i$)					& 0.3	& 0.9	& 0.2 \\ \hline
\enddata
\end{deluxetable}

Calculating a grid of model for the magnetospheric parameters requires the mass, radius, and effective temperature of the accreting object. These parameters are poorly constrained for PDS~70 planets, especially the recently discovered PDS~70c \citep{haffert2019}. For PDS~70b, the estimated mass range is as wide as ${\rm 2-17\,M_{Jup}}$ \citep{muller2018}, while the radius could be as small as ${\rm \sim1.3\,R_{Jup}}$, with an effective temperature of ${\rm \sim1200\,K}$ \citep{keppler2018}.
A similarly wide mass range, but slightly lower masses, has been given for PDS~70c \citep{haffert2019}, although no estimate of the radius is available. The effective temperature of PDS~70c is currently unconstrained, but there is a possibility that it is lower than that of PDS~70b since it has a redder spectrum \citep{haffert2019}.  Given the uncertainty in the planet parameters, we adopt a set of parameters that is reasonable for both planets for our exploration of the magnetospheric model. Therefore, we adopted a planet mass of ${\rm M_{p}=6\,M_{Jup}}$, a radius ${\rm R_{p}=1.3\,R_{Jup}}$, and an effective temperature ${\rm T_{eff}=1200\,K}$. These values are also applicable to young Jupiter-mass planets in general. 

We calculated a large grid of models with parameters covering the ranges shown in Table~\ref{tab:model_param}. In particular, the geometry of the magnetosphere covers the values predicted by \citet{batygin2018}, and the mass accretion rate spans 6 orders of magnitude covering the values calculated for PDS~70b and c \citep{haffert2019,wagner2018}. The temperature in the flow T$_{\rm max}$ covers the range found to be relevant for T Tauri stars at low accretion rate \citep{thanathibodee2019}.

\section{Results}\label{sec:results}
\subsection{Comparing the Models with Observations}

\begin{figure*}[t]
\epsscale{1.1}
\plotone{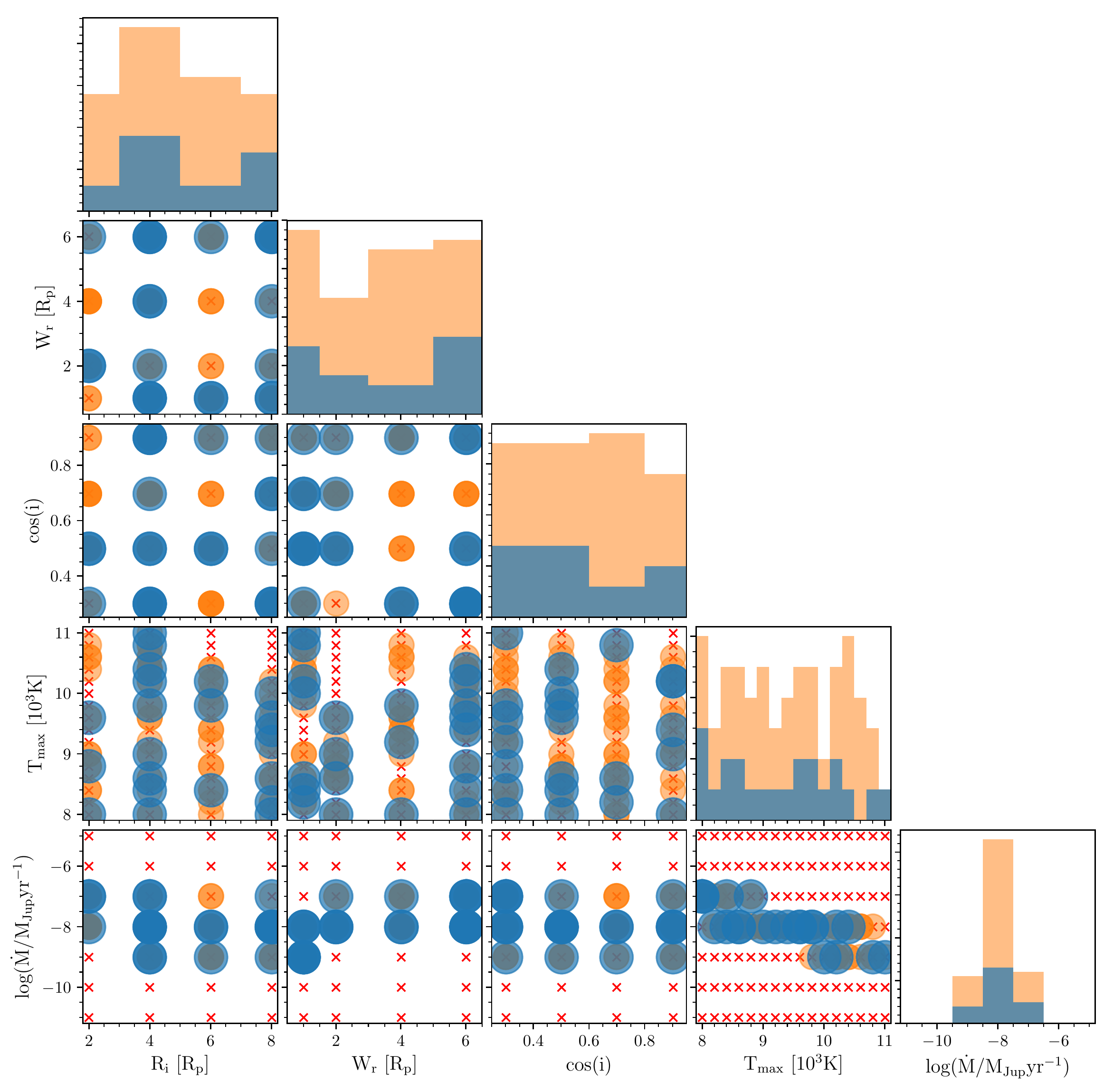}
\caption{Corner plot showing regions in parameter space that are in agreement with the observed {\halpha} flux of PDS~70b. Blue and orange points show the models which {\halpha} line flux are within 1$\sigma$ and 3$\sigma$ of the measured line flux of the planet, respectively. Red crosses show models outside of 3$\sigma$. The histograms show the distribution of the parameters of the models that fit
the observations, with the same color scheme as in scattered plots.
\label{fig:70b}}
\end{figure*}

\begin{figure*}[t]
\epsscale{1.1}
\plotone{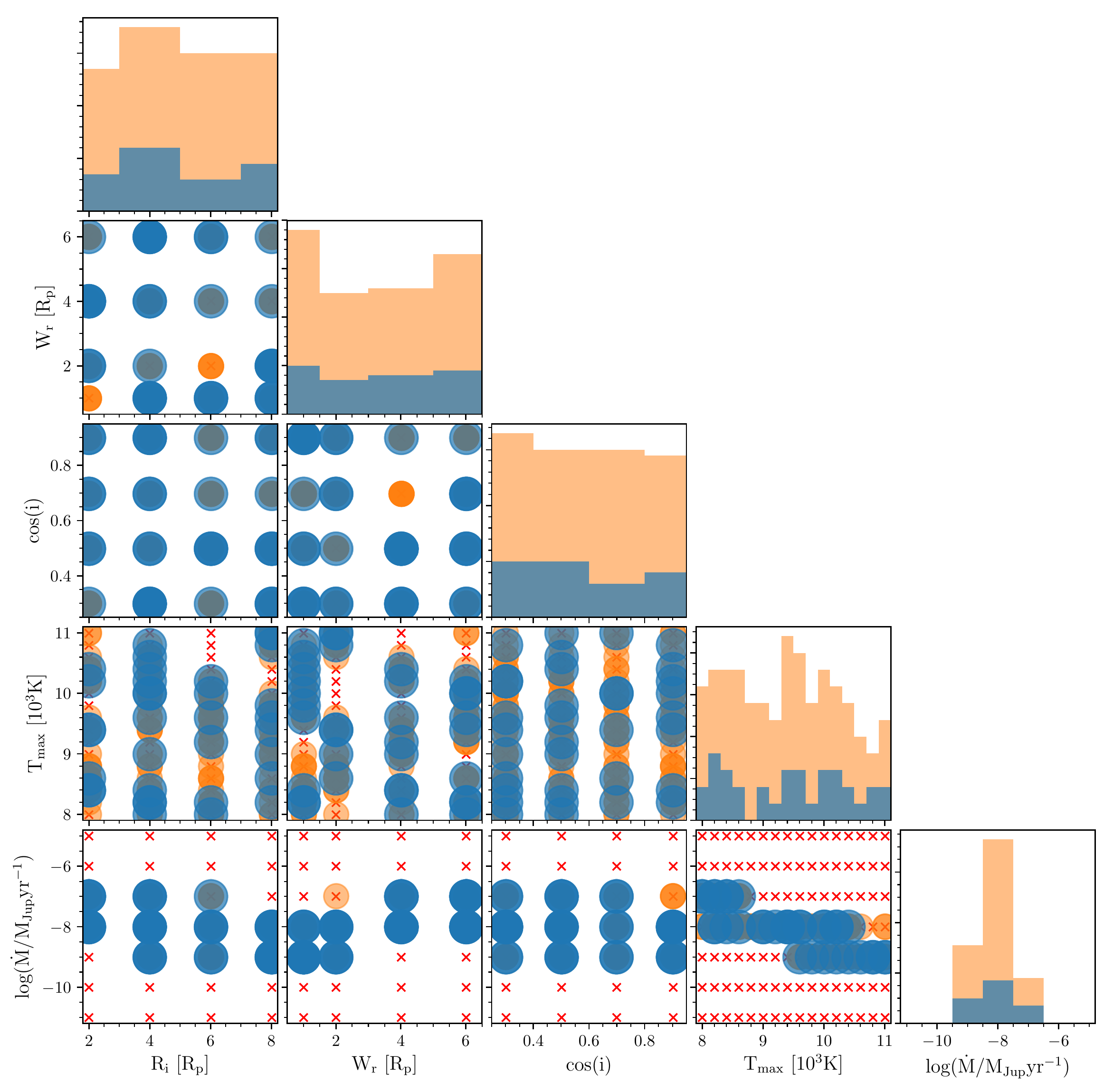}
\caption{Same as Fig.~\ref{fig:70b} but for PDS~70c.
\label{fig:70c}}
\end{figure*}

We compared the measured {\halpha} line fluxes of PDS~70b and PDS~70c with the line fluxes calculated in our grid of models. We did not take into account any extinction due to the disk \citep{wagner2018}, 
and first considered the measurements of \citet{haffert2019} as the intrinsic line fluxes 
for a consistent comparison between the two planets. Figures \ref{fig:70b} and \ref{fig:70c} show the scattered plots and histograms for combinations of the model parameters.
For the models with {\halpha} flux within the measurement uncertainties, the plots show flat distributions for \ri, \rw, and \cosi, suggesting that the uncertainty in geometric parameters have little effect on the final results. On the other hand, {\mdot} has a narrow distribution, implying that it is the strongest predictor for the line flux. Nevertheless, there is a degeneracy between {\tmax} and {\mdot}; similar line fluxes can be reproduced in high-T with low-{\mdot} and low-T with high-{\mdot}. This degeneracy has been noted in the original paper on the magnetospheric accretion model for T Tauri stars \citep{muzerolle2001}.

We calculated the mass accretion rates onto PDS~70b and c based on the models that predict {\halpha} line fluxes within 3$\sigma$ of the observed flux. By calculating the mean and standard deviation of the model log({\mdot}) we found log(\mdot) = $-8.0\pm0.6$\, $\mjupyr$ and log(\mdot) = $-8.1\pm0.6$\, $\mjupyr$ for PDS~70b and c, respectively. 
Although our mass accretion rate estimates are in agreement with those calculated by \citet{haffert2019} within uncertainties, we caution that this agreement is not because {\wten}-{\mdot} relation based on TTS \citep{natta2004} is applicable to accreting protoplanets (\S \ref{subsec:relationships}), but because {\wten} from VLT/MUSE measurements are overestimated due to the low spectral resolution. Given the planet's escape velocity of $\sim127\,\kms$, the intrinsic line width would be on the same order, and it is much narrower than the instrumental broadened observed line width with a spectral resolution $\sim110\,\kms$ at 6562\,\AA.

\begin{figure}[t]
\epsscale{1.2}
\plotone{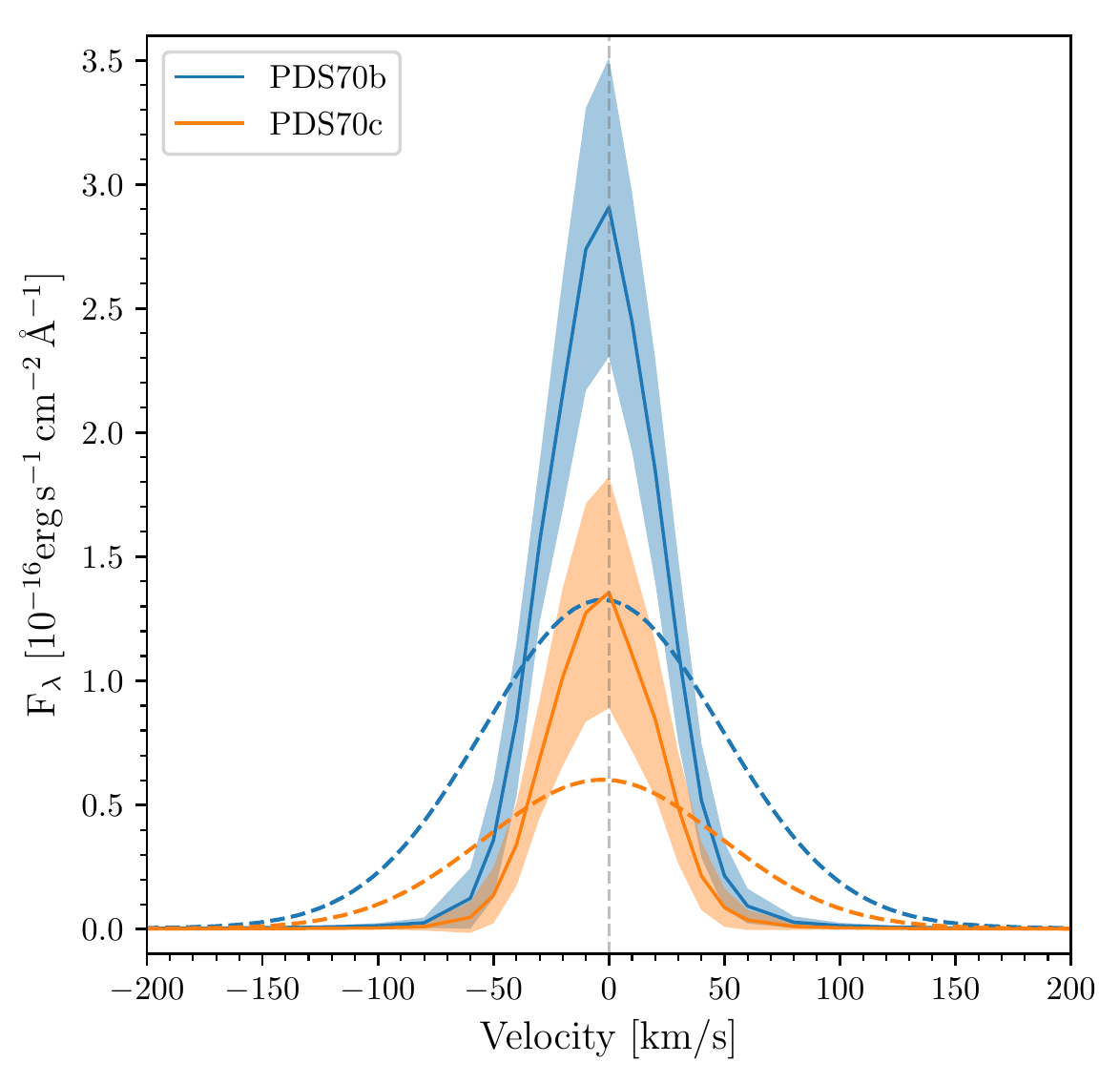}
\caption{{\halpha} line profiles of models that predicted line flux within 3$\sigma$ of the observed flux. The solid lines are the mean intrinsic line profiles from the models; shaded areas show standard deviations of the profiles. The dashed lines are the mean model line profiles convolved with a Gaussian instrumental profile to ${\rm R=2800}$, comparable to the resolution of MUSE at {\halpha} wavelength.
\label{fig:profiles}}
\end{figure}

As an illustration, Figure~\ref{fig:profiles} shows the mean {\halpha} line profiles from models with line flux within 3$\sigma$ of the observed flux. We measured the {\wten} of the mean model profile to be $100\,\kms$ and $96\,\kms$ for PDS~70b and c, respectively. On the other hand, the {\wten} for the profiles taking into account the spectral resolution of MUSE (R$\sim$2800at 6560\,\AA), are $224\,\kms$ and $219\,\kms$ for the two planets. Although these values are in agreement with \citet{haffert2019} measurements, they suggest that the instrumental broadening contributes significantly to the line width.

Another discrepancy between the results of our model and the observed {\halpha} profiles of PDS~70b and c is that the model does not show a line-center redshift, while \citet{haffert2019} reported a redshift of $25\pm8\,\kms$ and $30\pm9\,\kms$ for PDS 70b and c, respectively. One possibility is that this apparent line shift relative to the stellar {\halpha} line center is due to a shift of the line center of the star itself. The redshifted absorption component in the stellar {\halpha} line \citep{haffert2019} could cause the measured line center to be blueshifted. As a result, the line velocity measurement of the planet using the stellar {\halpha} line as a reference would appear redshifted. 
Higher resolution observations are required to test this suggestion.

We also compare the mass accretion rates inferred from the {\halpha} line fluxes of PDS~70b from MUSE with the \citet{wagner2018} determination using MagAO.
The value of the H$\alpha$ flux was not explicitly reported by \citet{wagner2018}, so
we followed the method outlined in \citet{close2014} using the H$\alpha$ contrast
and found an H$\alpha$ flux of ${\rm (3.3\pm1.8)\times10^{-15}\,erg\,s^{-1}\,cm^{-2}}$ for PDS~70b. 
We assume no extinction for a consistent comparison with the results of \citet{haffert2019}. 
The {\halpha} flux value from MagAO is an order of magnitude higher than that determined from the MUSE observations, which could imply that accretion onto PDS~70b could be variable. Using the method outlined above for models with fluxes within 1$\sigma$ of the MagAO flux, we found that the mass accretion rate of PDS~70b 
at the epoch of the MagAO observation was
log(\mdot) = $-7.8\pm0.7$\, $\mjupyr$. This 
value of
{\mdot} is similar to that based on the MUSE observation \citep{haffert2019}, mainly because the line flux is a steep function of {\mdot} (c.f. Fig.~\ref{fig:indicators}). 
Although the observed variability of the {\halpha} line flux could be due to variable accretion rate or extinction, detecting accretion variability is still challenging with the current assumptions of the model.
Detailed modeling that is more sensitive to accretion variability of the planet is a subject of a future study.

\subsection{Accretion Indicators for Planetary Mass Objects}
\label{subsec:relationships}
\begin{figure*}[t]
\epsscale{1.1}
\plotone{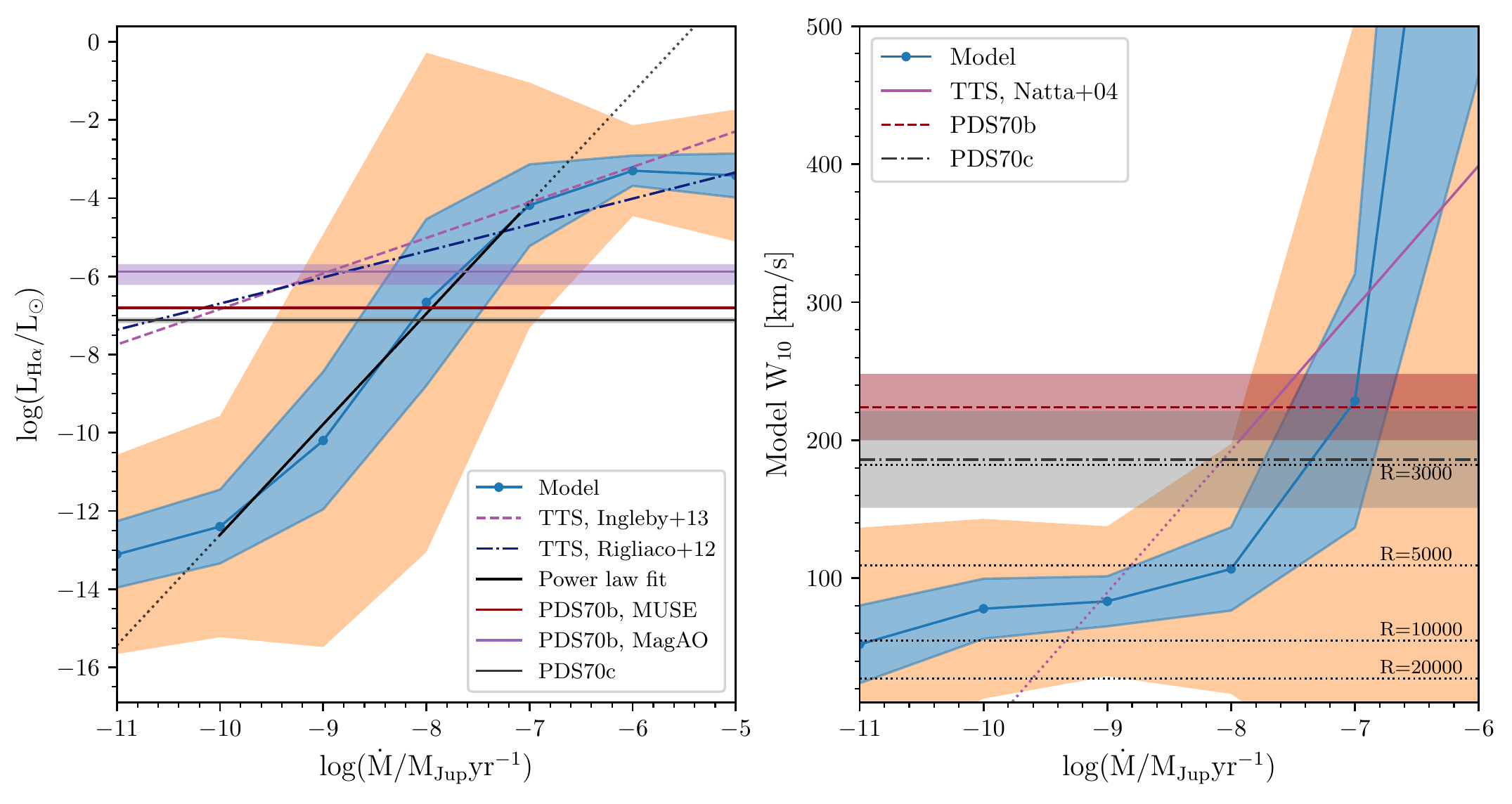}
\caption{
\emph{Left:} Predicted {\halpha} luminosity as a function of mass accretion rate for planets with parameters similar to those of PDS~70b and c. The green and pink shaded regions show 1$\sigma$ and 3$\sigma$, respectively, around the mean value (blue line) for a given {\mdot}. The line luminosity saturates once the accretion rate reaches the low end of mass accretion rate in pre-main sequence stars (T Tauri stars, TTS), 
{\mdot}\,$\sim10^{-10}\,\msunyr\sim10^{-7}\,\mjupyr$. 
The dashed purple line and the dash-dotted dark-blue line are extrapolation of empirical relationships between {\mdot} and {\lhalpha} for TTS. The relationships underestimate the {\mdot} of the planets by $\sim$2 orders of magnitude. The solid black line is the power law fit to the relationship for log(\mdot) between -10 and -7; the dotted line is the extrapolation of the relationship outside the fitted range. Shaded horizontal lines show the line measurements of PDS 70 planets.
\emph{Right:} The {\wten} of the model {\halpha} line as a function of log(\mdot). The purple lines are the relationship for T Tauri stars from \citet{natta2004}, with extrapolation to {\wten\,$<200\,\kms$} in dotted line. The line width is a very weak function of {\mdot}. Horizontal dotted lines show {\wten} of Gaussian instrumental profiles for several spectral resolutions. The shaded crimson dashed line and dash-dotted gray line show the measured {\wten} of PDS~70b and PDS~70c, respectively.
\label{fig:indicators}}
\end{figure*}

Empirical relationships between {\halpha} line luminosity (\lhalpha), or 
{\wten} of the line, and {\mdot} have been extensively used for T Tauri stars \citep{rigliaco2012,ingleby2013,natta2004}, and they have been adopted to estimate the {\mdot} of accreting planets \citep{wagner2018,haffert2019}. To examine the validity of those relationships in the 
planetary regime,
we calculated the mean and standard deviation of {\lhalpha} and {\wten} for a given value of {\mdot} in our model grid, shown in Figure~\ref{fig:indicators}. The shaded regions correspond to 1$\sigma$ and 3$\sigma$ from the mean. As expected, {\lhalpha} increases as {\mdot} increases. 
The relationship between {\mdot} and {\lhalpha} 
can be represented by a power law
between $\sim10^{-10}$ to $\sim10^{-7}\,\mjupyr$, and the line flux seems to
saturate at the high end of the accretion rate (\mdot$\sim10^{-6}\,\mjupyr$); this trend has also been found in TTS \citep{ingleby2013}. 
At the low {\mdot} end, the relationship became nearly flat, which could be due to
the weakness of the line at low density, essentially rendering it undetectable against the continuum.

Using simple linear regression, we fitted a power law 
to the {\lhalpha} model results between {\mdot} = $10^{-10}-10^{-7}\,\mjupyr$, 
yielding the relationship
\begin{equation}
    \log({\rm L_{H\alpha}}) = (2.83\pm0.02)\log(\dot{\rm M}) + (15.7\pm0.2), \label{eq:mdot-lum}
\end{equation}
where {\lhalpha} is in solar luminosities and {\mdot} in $\mjupyr$. To aid the interpretation of future 
observations
of
planets with similar parameters, we also fitted the inverse of this correlation, treating {\lhalpha} as an independent variable, which yields
\begin{equation}
    \log(\dot{\rm M}) = (0.280\pm0.002)\log({\rm L_{H\alpha}}) - (6.14\pm0.02).
\end{equation}
This relationship is different than the inverse of Eq.~\ref{eq:mdot-lum} since the dependent variable (i.e., {\mdot}) is not fully sampled.

As a comparison, in Figure~\ref{fig:indicators} we plot 
the relationships between {\mdot} and L$_{\rm H\alpha}$ from \citet{ingleby2013}
and \citet{rigliaco2012}, the latter derived
from the correlation between
L$_{\rm acc}$ and {\lhalpha}
using the parameters of the planets.
The \citet{ingleby2013} relationship is calculated from observations of a large number of T Tauri stars across a wide range of ages using contemporaneous UV and optical spectra, providing direct measurements of the accretion luminosity from the UV and of the line luminosity from the optical. Similarly, \citet{rigliaco2012} used simultaneous observations, but
for a lower mass range. As shown in Figure~\ref{fig:indicators}, the relationship found in this study is steeper than those 
for T Tauri stars
with up to two orders of magnitude difference in {\mdot} for the parameters of the PDS~70 planets. 
The relationships are clearly different,
even considering
the intrinsic uncertainty in our modeling results, which as shown by the blue shaded region 
in Figure~\ref{fig:indicators}, can be as high as one order of magnitude in {\mdot}.

The right panel of Figure~\ref{fig:indicators} shows the relationship between {\mdot} and {\wten} for all models in our grid.
At low accretion rates, 
{\mdot} 
$ < 10^{-8}\,\mjupyr$, the line width is not sensitive to {\mdot} even 
if it varies by orders of magnitude.
In contrast, {\wten} increases steeply with {\mdot} for {\mdot} $\gtrsim10^{-8}\,\mjupyr$ with a scatter significantly increased.
This happens because even though the mass accretion rates are comparable only to
those of the lowest accretors among the TTS, with
smaller spatial scales  the density
in the accretion flow becomes high 
enough for the line wings to become 
optically thick and  
pressure broadening 
to become effective \citep{muzerolle2001}.  
The steep rise also reflects the high temperature in the model; 
{\tmax} is found to be $\sim7000-8000$\,K for 
TTS magnetospheres in this range of {\mdot} \citep{muzerolle2001}, whereas our models
have higher values of {\tmax}, which are applicable at low {\mdot} \citep{thanathibodee2019}. 
The large scatter of {\wten}
at high {\mdot} is due to the treatment of the continuum optical depth in the model. 
For the high-density cases, the continuum opacity of the magnetosphere 
can become so high that the
continuum arises in the flow, 
and the photosphere of the planet is hidden.
In this case,
the {\halpha} line emerging from the flow becomes very weak compared to the new continuum, and in some cases, the line appears in absorption.
As a result, the measurements of {\wten} in this regime become meaningless since the continuum is now uncertain.

Plotted in purple line in the right panel of Figure~\ref{fig:indicators} is 
the empirical relationship between {\wten} and {\mdot} for T Tauri stars from \citet{natta2004}. 
Again, the relationship for T Tauri stars is different from our model results for planets. One possibility for this discrepancy is that 
the \citet{natta2004} relationship is based on measurements
of TTS  covering a broad range of masses and radii, while our relationship is only applicable for a specific mass and radius. 
It is conceivable that one could fine-tune the model such that it reproduces the relationship, at least in the high {\mdot} regime that is well-calibrated in TTS. 
Such a study is beyond the scope of this paper, and the results would not be meaningful without calibrations with more planetary-mass accretors at their applicable accretion rates. 
At the moment, such a study is still challenging with
the low-resolution 
of the 
existing planet spectrum. 
For example, the figure also shows the {\wten} of a purely Gaussian instrumental profile for different spectral resolutions. 
In order to extract geometric information from the line, sufficiently high signal-to-noise at the line wing is required. 
High-resolution spectra, comparable to the line width, will help to accurately measure {\wten} by suppressing the background noise from the star. 
With more detected planetary spectra, the relationship between {\mdot} and {\wten} in the planetary mass regime can be revisited in the future.

\section{Discussion and Conclusion} \label{sec:discussion}

Following the detection of two accreting planets around the K7
star PDS~70, based in part on the 
measurement
of {\halpha} line emission from the planets \citep{wagner2018,haffert2019}, we 
applied 
the magnetospheric accretion model \citep{muzerolle2001}
to the material accreting onto the
planet to test if 
the model could explain the observed line flux with a reasonable range of parameters. 
Our study is the first to apply the full treatment of {\halpha} line radiative transfer in 
a magnetospheric geometry for planetary-mass objects. 
The model can reproduce the observed fluxes, indicating that a magnetospheric model comparable to that of TTS
can still be applicable in the planetary mass regime, as suggested by \citet{batygin2018}. 
Our results also suggested that accretion can be the main source of {\halpha} line emission from planets in circumstellar disks, and confirmed the use of {\halpha} as a viable method to detect young planets.

Measuring the mass accretion rate of a planet from the
flux of the {\halpha} emission line is still challenging due to practical considerations, such as the low resolution of the IFU spectrographs used to detect the line,
as well as the uncertainty in the line luminosity due to background subtraction and extinction \citep{wagner2018,haffert2019}. 
An additional challenge comes from the
uncertain calibration between the 
{\halpha} luminosity
and the accretion rate
in this very low-mass regime. 
Our results provide, for the first time, 
a correlation between the {\halpha} line luminosity and {\mdot}
based on a physical model. In addition, we show
that this correlation
is different from
the relationships applicable to
T Tauri stars. 
However, we note the limitations of using our result
since it requires knowledge of the planetary parameters, i.e., mass, radius, and effective temperature, 
from independent measurements. 
Large uncertainties in these values can change the final measured accretion rate in an uncertain way since the line emission properties depend on the dynamic of the
gas in the accretion flow, which in turn depends on the mass of the object and the physical size of accretion flow; these also determine the optical depth of the line. 
Our provided relationship between {\lhalpha} and {\mdot} is therefore only applicable for planets with 
mass and radius comparable to those of PDS~70b and c. 
Similar limitations apply to the calibrations of {\mdot} with {\wten} since the line width is more sensitive to the planet parameters.
The applicability of the magnetospheric accretion model in different ranges of 
planetary parameters will be the subject of future studies.

As in TTS, the highest uncertainty in measuring the mass accretion rate 
arises from the unknown temperature in the flow
\citep{muzerolle2001}.
Magnetohydrodynamic simulations with a full treatment of magnetic and other heating sources in the flow are still needed to shed light onto the temperature problem. Observations of other hydrogen line transitions and other line species may also help to constrain the temperature empirically.

From an observational standpoint, modeling the velocity-resolved non-Gaussian {\halpha} line profiles using the magnetospheric model \citep{muzerolle2001,thanathibodee2019} will provide much more accurate measurements of the mass accretion rates, since both the line flux and the line width can be simultaneously constrained. 
Such observations are still unavailable due to limitations of
the current technology. High-resolution IFU spectrograph on a large telescope, such as HARMONI on the ELT will provide data necessary for such modeling study.

\acknowledgments
\noindent
This project is supported in part by NASA grant NNX17AE57G.
We thank Lee Hartmann and Michael Meyer for insightful suggestions.

\bibliography{2019b}{}
\bibliographystyle{aasjournal}

\end{document}